\documentclass[twocolumn,prl,preprintnumbers,amsmath,amssymb]{revtex4}

\usepackage{graphicx}
\usepackage{dcolumn}
\usepackage{bm}

\begin{document}

\title{A Superconducting instability in the surface of a topological insulator}

\author{Alberto Cortijo}
\affiliation{Department of Physics, Lancaster University,
Lancaster, LA1 4YB, United Kingdom}%
\affiliation{Departamento de F\'{i}sica te\'{o}rica, Universidad Aut\'{o}noma de Madrid, E-28049, Madrid, Spain}%


\begin{abstract}
It is shown that a superconducting instability appears in the electronic states on the surface of a topological insulator due purely to electromagnetic interactions. The discussion of this instability is based on the analysis of the gap equation using the repulsive term coming from the axion term and the assumption that the system behaves like a two dimensional Fermi liquid. It is shown that this superconducting instability appears in the p-wave channel and an estimate of the critical temperature of the transition is given.
\end{abstract}

\maketitle
\emph{Introduction.} The topological insulating (TI) state offers the possibility to study physical phenomena that were considered as exclusive of the High Energy Physics not so long ago\cite{Hsieh08,Hsieh09,Xia09}. The prominent example is the topological magnetoelectric coupling known in the former context as the \emph{axion} term \cite{Qi08,Wilczek87}. This term has proved to be quite attractive to the Condensed Matter community due to its fascinating properties like the possibility of the existence of Majorana Fermions in superconductor-TI junctions and a repulsive Casimir interaction between topological insulating plates\cite{Fu08,Grushin11} (or an up-to-date review of the subject, see \cite{Kane10}). Another interesting example, a superconducting state on the surface of a three dimensional TI, has been analyzed in a context of a general symmetry analysis or in basis of a generic short range interactions\cite{Santos10,Fu10}. From the experimental side, a superconducting phase has been observed Bi$_{2}$Se$_{3}$ doped with Cu\cite{Hor10,Wray10}. It is observed that the presence of a Cu layer is the key ingredient to induce superconductivity. Here our aim is not to explain these experimental observations but rather suggest a microscopical mechanism of superconductivity when the TI surface states are magnetically gaped being the magnetoelectric term the ultimate responsible for the electronic pairing. The problem is of considerable interest for several reasons: it constitutes a physical realization of the residual pairing mechanism proposed to occur in the context of the Quantum Hall Effect (QHE) at half filling\cite{Greiter92}, and also the resulting superconducting order parameter enjoys p-wave symmetry opening the possibility of finding majorana states at the vortices in this system, in this case without relying in any proximity effect\cite{Ivanov01,Fu08}.


\emph{The model.} Our starting point is the action for the electronic states coupled to the electromagnetic field:

\begin{equation}
S=\int d^{4}x\left(\delta(x_{3})\mathcal{L}_{surf}+\mathcal{L}_{EM}+\mathcal{L}_{bulk}\right).\label{totalaction}
\end{equation}

We will assume that while the surface electronic states live on the surface of the TI (this is the origin of the term $\delta(x_{3})$) both the bulk gapped electronic states \emph{and} the electromagnetic field live in the three spatial dimensions. This fact implies that we can not longer describe the electrodynamics of the surface states by a proper QED$_{3}$, as is done in \cite{Kogan89,Girotti92} and some crucial differences arise when the photon propagator is calculated, as we will see. Another important difference is that although the spectrum of the electronic states is well described by a Dirac equation, the typical Fermi velocity of these states is around c/600\cite{Zhang09} and then we expect that retardation effects are negligible. We will use $\beta\equiv v_{F}/c$ as a perturbation parameter to keep terms that will lead to the effective instantaneous Coulomb potential.

First of all, we will focus on the electromagnetic part of the action in (\ref{totalaction}). Because we are mainly interested in the non retarded effects of the Coulomb interaction in the surface states, we will not explicitly write the term corresponding to the gapped bulk states. Instead of that, we will integrate out the bulk fermions taking them into account in the modification induced in the action for the electromagnetic action. It is already well known that this contribution is twofold: on the one hand, we consider the system to be a nonmagnetic insulator ($\mu\simeq 1$), that is, the effects of the bulk electronic structure (and other effects at higher frequency scales) are encoded in the dielectric susceptibility $\epsilon(\omega,\mathbf{k})$. Moreover, here we are dealing with physical effects at a very low energy scale, smaller than the gap energy of the bulk electrons and other polarization effects so it is reasonable to approximate $\epsilon$ by its value at zero frequencies and momenta: $\epsilon(\omega,\mathbf{k})\sim \epsilon(0,0)\equiv \epsilon_{r}$. On the other hand, an axion-like term also arises in the electromagnetic response of the bulk electrons in a TI\cite{Qi08}. In order to make the axion term physically observable we need to introduce a time reversal breaking element in the surface of the TI, like a magnetic field, a ferromagnetic coating or magnetic impurities. The side effect of such an element is to open a gap in the surface electronic spectrum\cite{Chen10}.
This term, called magnetoelectric term, strongly modifies the Maxwell equations in the material medium\cite{Wilczek87} and thus it becomes evident that if we want to consider the electromagnetic field propagating in the three spatial dimensions we must take into account the presence of such a term and thus it will modify the long range Coulomb interaction among electrons.

We will follow the strategy settled in \cite{Gonzalez94} to compute the EM propagator. Because the electromagnetic field is coupled to a a current constrained to live on a surface, we can get rid of the fourth component of the electromagnetic field $A_{\mu}(x)$ by using a suitable choice of gauge. In this context we choose the Feynman gauge $\partial^{\mu}A_{\mu}=0$ because this particular choice of gauge not only allows us to decouple the dynamical evolution of $A_{3}$ from the other components but it makes the even part of the photon propagator to be proportional to the surface metric tensor $\eta_{ij}=diag(1,-1,-1)$.

Before discussing the action of the electromagnetic field $A_{j}(x)$ we can take some advantage of the fact that we are considering a constant dielectric function $\epsilon_{r}$. We anticipate that the effective Coulomb interaction at lowest order in perturbation theory in the charge $e$ will be of the order of $e^{2}$. Also we know that the dielectric constant always appears in the form of the ratio $e^{2}/\epsilon_{r}\equiv\alpha_{r}$ so we can safely work with a free electromagnetic action throughout the intermediate calculations and then just substitute in the final result the bare charge by this ratio.

Under these considerations, the effective action for the electromagnetic field including the axion term will be:

\begin{equation}
S_{EM}=\int d^{4}x\left(A_{\mu}\eta^{\mu\nu}\partial^{2}A_{\nu}+\alpha\theta(x)\epsilon^{\mu\nu\rho\lambda}\partial_{\mu}A_{\nu}\partial_{\rho}A_{\lambda}\right),\label{emaction}
\end{equation}
where $\alpha$ is the fine structure constant and $\theta(x)$ is the axion field. We can model the axion field for a perfect interface vacuum-TI with a step function in the z direction: $\theta(x)=\theta\Theta(x_{3})$ where now $\theta$ takes the value of $(2n+1)\pi$, $n\in \mathbb{N}$. Integrating by parts the second term in (\ref{emaction})  and integrating out $A_{3}$, $S_{EM}$ becomes

\begin{equation}
S_{EM}=\int d^{4}x (A_{i}\eta^{ij}\partial^{2}A_{j}-\frac{\alpha\theta}{2\pi} \delta(x_{3})\hat{n}_{3}\epsilon^{3ijk}A_{i}\partial_{j}A_{k}).\label{emaction2}
\end{equation}
In (\ref{emaction2}) we see that the axion term acquires the form of a Chern-Simons (CS) term\cite{Qi08}. However the dependence of the electromagnetic field of the coordinate $x_{3}$ radically alters the form of the electromagnetic propagator $D_{ij}(x,x')$ and it will be completely different to what one expects from the electromagnetic propagator in a CS QED$_{3}$\cite{Girotti92}. The first thing we have to notice is that now the system is no longer translationally invariant along the direction $x_{3}$ so the spatial dependence of the propagator will be $D_{ij}(\mathbf{x}-\mathbf{x}',x_{3},x'_{3})$. Because now the electromagnetic field is coupled to a system which lies on a surface, we do not need to know the complete dependence of $D_{ij}$ with $x_{3}$ and $x'_{3}$ but only the value at the surface of $D_{ij}(k,0,0)$ (the Fourier transform in the rest of the coordinates). By noticing that the second term in (\ref{emaction2}) is proportional to $\delta(x_{3})$ and using the Dyson equation for the electromagnetic propagator we get:

\begin{eqnarray}
D_{ij}(q,0,0)=\frac{1}{D_{0}^{ij-1}(q,0)+i\frac{\alpha\theta}{2\pi}\epsilon^{ijl}q_{l}}.\label{propagator1}
\end{eqnarray}
In the last expression, $D_{0}^{ij-1}(q,0)=D_{0}^{ij-1}(q,x_{3}-x'_{3}=0)$ stands for the integral of the free photon propagator which in momentum representation reads\cite{Gonzalez94}:

\begin{equation}
D^{0}_{ij}(q,0)=\int\frac{dq_{3}}{2\pi}D^{0}_{ij}(q,q_{3})=\frac{-i\eta_{ij}}{2|q|},\label{freepropagator}
\end{equation}
and $|q|=\sqrt{q^2_{0}-\mathbf{q}^2}$. Note that now both terms in the denominator of (\ref{propagator1}) are linear in $q$. Then using (\ref{propagator1}) and (\ref{freepropagator}) we arrive at the expression for $D_{ij}(q,0,0)$ that we will use to compute the effective Coulomb interaction (in the nonretarded limit $q_{0}\rightarrow 0$):

\begin{equation}
D_{ij}(\mathbf{q},0,0)=\frac{-1}{2\left(1+\frac{\alpha^2\theta^{2}}{4\pi^{2}}\right)}\frac{1}{|\mathbf{q}|}\left[\eta_{ij}-\frac{i\alpha\theta}{2\pi}\epsilon_{ijl}\frac{q^{l}}{\mathbf{|q|}}\right].\label{effpropagator}
\end{equation}
The effective propagator for the electromagnetic interaction is not equal to the propagator for a purely CS QED$_{3}$ theory. In the present case, the effective propagator is still massless, despite of the presence of a CS term, contrary to what happens when the photon only propagates in two spatial dimensions, where the CS term induces a mass for the photon.

Let us turn our attention to the action of the surface electronic states. In order to make the axionic term observable we have to break the time reversal symmetry in the system. This situation can be experimentally achieved by directly adding a magnetic coating on the surface or by adding enough magnetic impurities in the bulk of the TI\cite{Chen10}. The side effect is that we induce a mass in the surface electronic spectrum. In the action (\ref{totalaction}) we will define $x^{0}=ct$ so that the action term for the surface states is
\begin{equation}
S_{surf}=\int d^{4}x\delta(x_{3})\bar{\psi}\left(-i\gamma_{i}M^{ij}(\partial_{j}+ieA_{j})-m\right)\psi.\label{surfaction}
\end{equation}
In (\ref{surfaction}) the indices $(i,j)$ run only from $0$ to $2$ and we have defined the matrix $M^{ij}=diag(1,-\beta,-\beta)$. Since the parameter $\beta$ is already defined, we will set $c=1$ in the rest of the paper. The matrix convention used in (\ref{surfaction}) is $\gamma_{0}=\sigma_{z},\gamma_{1}=i\sigma_{y}$, and $\gamma_{2}=-i\sigma_{x}$, so the spin operator for the surface estates is $\vec{S}=\hat{z}\times \vec{\sigma}=-i\vec{\gamma}$. Going to a second quantization formalism, we can expand the field operators in terms of creation and annihilation operators of states in the valence ($b_{\mathbf{k}}$) and conduction bands ($a_{\mathbf{k}}$):
\begin{eqnarray}
\psi(\mathbf{x})&=&\sum_{\mathbf{k}}e^{i\mathbf{kx}}u_{\mathbf{k}}a_{\mathbf{k}}+e^{i\mathbf{-kx}}v^{*}_{\mathbf{k}}b^{+}_{\mathbf{k}},\nonumber
\\
\bar{\psi}(\mathbf{x})&=&\sum_{\mathbf{k}}e^{-i\mathbf{kx}}u^{*}_{\mathbf{k}}\gamma^{0}a^{+}_{\mathbf{k}}+e^{i\mathbf{kx}}v_{\mathbf{k}}\gamma^{0}b_{\mathbf{k}},\label{fieldoperators}
\end{eqnarray}
where $u_{\mathbf{k}}$ and $v_{\mathbf{k}}$ are the planewave solutions of the Dirac equation resulting from (\ref{surfaction}) ($v_{\mathbf{k}}=\gamma_{1}u^{*}_{\mathbf{k}}$):
\begin{eqnarray}
u_{\mathbf{k}} &=&\sqrt{\frac{m+\omega_{\mathbf{k}}}{2\omega_{\mathbf{k}}}}\left(\frac{\beta(k_{x}-ik_{y})}{m+\omega_{\mathbf{k}}},1\right)^{T}.\label{spinors}
\end{eqnarray}
In what follows we will set the Fermi level slightly above the bottom of the conduction band so in virtue of the smallness of $\beta$ we can write $\omega_{\mathbf{k}}-\mu\equiv\sqrt{\beta^2\mathbf{k}^2+m^2}-m\approx \beta^{2}\mathbf{k}^2/2m=\varepsilon_{\mathbf{k}}$.

According to (\ref{effpropagator}), the exchange of a photon induces an effective electron-electron interaction containing two terms corresponding to the even (proportional to $\eta_{ij}$) and odd (proportional to $\epsilon_{ijl}$) terms in the effective propagator. The even part gives rise to the standard non-retarded Coulomb interaction. As it happens in standard two dimensional electron gases the existence of a finite Fermi surface will induce a polarization screening of the Coulomb interaction and hence a quasiparticle lifetime proportional to $\omega^{2}\ln\omega$, for energies $\omega$ close enough to the Fermi level, so we can safely consider a Fermi liquid picture where the Coulomb repulsion among electrons leads to a weakly interacting system of fermionic quasiparticles with renormalized mass $m$ and Fermi velocity $\beta$ and where an extra interaction term, corresponding to the odd part of the effective propagator (\ref{effpropagator}) remains. Of course, if we insist in the static screening of the Coulomb interaction we are forced to accept that the odd part of the propagator also changes by polarization effects so we need to know how the bare odd term gets modified by the inclusion of the polarization term $\Pi_{ij}(\omega,\mathbf{q})$ in (\ref{effpropagator}) in the static, long wavelength limit. Both the even and odd parts of the polarization function $\Pi_{ij}(\omega,\mathbf{q})$ are well known in the literature \cite{Chaichian98}. Gauge invariance tell us that the polarization tensor is defined through two scalar functions, $\Pi^{even}(\omega,\mathbf{q})$ and $\Pi^{odd}(\omega,\mathbf{q})$. At zero frequency and in the long wavelength limit they take the values $\Pi^{even}(0,\mathbf{q}\rightarrow 0)\sim 2\mu/\pi\equiv\lambda$ ($\lambda$ is the Thomas-Fermi wavevector) and $\Pi^{odd}(0,\mathbf{q}\rightarrow 0)\sim m/\mu$ so we can write the following expression for the inverse of the odd part of the effective propagator ($\gamma\equiv\alpha_{r}(\theta+\epsilon_{r}\frac{m_{r}}{\mu})/2\pi$):
\begin{equation}
D^{odd}_{ij}(0,\mathbf{q})=-\frac{\gamma}{(|\mathbf{q}|+\lambda)^{2}+\gamma^{2}|\mathbf{q}|^2}\epsilon_{ijl}q^{l}.\label{effdressedpropagator}
\end{equation}

Because now we are interested in the regime of energies close to the Fermi level, we will only take into account terms in the interaction hamiltonian corresponding to intraband processes, so the relevant interaction term coming from the odd part of the dressed effective propagator reads in the Cooper channel $\mathbf{q}=\mathbf{k'-k}$:

\begin{equation}
H_{int}=\sum_{\mathbf{k},\mathbf{k'}}V(\mathbf{k,k'})a^{+}_{\mathbf{k}}a_{\mathbf{k'}}a^{+}_{\mathbf{-k}}a_{\mathbf{-k'}},\label{effhamiltonian}
\end{equation}

where the potential $V(\mathbf{k,k'})$ is defined through the odd part of the effective propagator, the electron wavefunctions $u_{\mathbf{k}}$ and the matrix $M^{ij}$:
\begin{equation}
V(\mathbf{k,k'})=\alpha_{r}D^{odd}_{ij}(0,\mathbf{k'-k})M^{ir}M^{js}u^{*}_{\mathbf{k}}\gamma_{r}u_{\mathbf{k'}}u^{*}_{\mathbf{-k}}\gamma_{s}u_{\mathbf{-k'}}.\label{effpotential}
\end{equation}

In order to find the form of $V(\mathbf{k,k'})$ we need to know the form of the vector $\Gamma_{i}(\mathbf{k},\mathbf{k'})\equiv u^{*}_{\mathbf{k}}\gamma_{i}u_{\mathbf{k'}}$. Using the explicit form of $u_{\mathbf{k}}$ in (\ref{spinors}) up to order $\beta^{2}$ we have
\begin{eqnarray}
\Gamma_{0}(\mathbf{k},\mathbf{k'})=1-\frac{\beta^{2}}{4m}\left(|\mathbf{k-k'}|^2+i\hat{z}(\mathbf{k'}\times\mathbf{k})\right),\nonumber
\\
\Gamma_{1}(\mathbf{k},\mathbf{k'})=\frac{\beta}{2m}(k'+k^{*}),
\Gamma_{2}(\mathbf{k},\mathbf{k'})=\frac{-i\beta}{2m}(k'-k^{*}),\label{gammavectors}
\end{eqnarray}
where $k=k_{x}+ik_{y}$ and $k^{*}=k_{x}-ik_{y}$. It is worth to mention that the zero component of $\Gamma_{i}$ is even under the simultaneous change $\mathbf{k},\mathbf{k'}\rightarrow-\mathbf{k},-\mathbf{k'}$ while the spatial components $\Gamma_{(1,2)}$ are odd. This fact prevents the summation in (\ref{effpotential}) to be zero.

With all these considerations in mind, we arrive to the explicit form of the effective potential (\ref{effpotential}):

\begin{equation}
V(\mathbf{k,k'})=-\frac{\alpha_{r}\gamma\beta^{2}}{m}\frac{|\mathbf{k-k'}|^{2}-2i\hat{z}(\mathbf{k'}\times\mathbf{k})}{(|\mathbf{k-k'}|+\lambda)^{2}+\gamma^{2}|\mathbf{k-k'}|^{2}}.\label{effpotential2}
\end{equation}
The equation for the gap $\Delta_{\mathbf{k}}$ can be easily written by using an standard Bogoliubov transformation and taking into account (\ref{effhamiltonian}) and (\ref{effpotential2}) :

\begin{equation}
\Delta_{\mathbf{k}}=\frac{1}{2}\int \frac{d^{2}\mathbf{k'}}{4\pi^2}\frac{\Delta_{\mathbf{k'}}}{\sqrt{\Delta_{\mathbf{k'}}^2+\varepsilon^{2}_{\mathbf{k'}}}}V(\mathbf{k,k'}).\label{gapequation}
\end{equation}

\emph{Results.} The presence of the complex unity in the second term in (\ref{effpotential2}) prevents $\Delta_{\mathbf{k}}$ to enjoy s-wave symmetry. This is expected, however because the time reversal symmetry is broken due to the presence of the magnetic coating, responsible of the mass of the surface electrons. We will employ a p-wave symmetric ansatz for the gap parameter: $\Delta_{\mathbf{k}}=\Delta(k_{x}+ik_{y})=\Delta|\mathbf{k}| e^{i\phi_{\mathbf{k}}}$. Also, the Thomas-Fermi wavevector $\lambda$ introduces a characteristic lengthscale in the problem. Because of that, two different regimes in eq. (\ref{gapequation}) allows for analytical tractability: $|\mathbf{k}|,|\mathbf{k'}|\ll\lambda$ and $|\mathbf{k}|,|\mathbf{k'}|\gg\lambda$, leading to upper and lower bounds of the transition temperature $T_{c}$ respectively.

In the regime $|\mathbf{k}|,|\mathbf{k'}|\ll\lambda$ we can substitute the denominator in (\ref{effpotential2}) simply by $\lambda^{2}$. By the shift $\phi_{\mathbf{k'}}\rightarrow\phi_{\mathbf{k'}}-\phi_{\mathbf{k}}$ the angular integration in (\ref{gapequation}) is straightforward leading to:

\begin{equation}
\Delta=\frac{\alpha_{r}\gamma\beta^{2}}{8\pi^{2}\lambda^{2}}\int^{\Lambda}_{0} d|\mathbf{k'}|\frac{\Delta|\mathbf{k'}|^{2}}{\sqrt{\Delta^{2}+\frac{\beta^{4}}{4m^2}|\mathbf{k'}|^2}}.\label{gapeq1}
\end{equation}
Because we are formally working in the limit $\lambda\rightarrow\infty$ we can substitute the upper limit of the $|\mathbf{k}|$ integral $\Lambda$ by the Thomas-Fermi wavevector $\lambda$ and keep the leading term in the limit $\beta\rightarrow0$. The result is simply
\begin{equation}
\Delta=\frac{\alpha_{r}\lambda\gamma\beta^{2}}{6\pi}.
\end{equation}
In the most favorable situation where $\beta=1/600$, $\mu\sim m\approx7$meV, $\theta=\pi$, and $\epsilon_{r}\sim1$, the critical temperature is estimated in this limit to be $T_{c}\sim\frac{\Delta}{k_{B}}\approx 6$K.

In the opposite regime $|\mathbf{k}|,|\mathbf{k'}|\gg\lambda$, the lengthscale $\lambda$ disappears from (\ref{effpotential2}). As we have mentioned in the introduction it is interesting to note that such limit strongly resembles to the attractive residual interaction among quasiparticles in the half filling regime in the QHE\cite{Greiter92}. After performing the angular integration, eq. (\ref{gapequation}) takes the form

\begin{eqnarray}
\frac{4\pi m}{\beta^{2}\gamma}\Delta|\mathbf{k}|=\int^{|\mathbf{k}|}_{0} d|\mathbf{k'}|\frac{\Delta|\mathbf{k'}|}{\sqrt{\Delta^{2}+\frac{\beta^{4}}{4m^{2}}|\mathbf{k'}|^{2}}}\frac{|\mathbf{k'}|}{|\mathbf{k}|}+\nonumber
\\
+\int^{\infty}_{|\mathbf{k}|} d|\mathbf{k'}|\frac{\Delta|\mathbf{k'}|}{\sqrt{\Delta^{2}+\frac{\beta^{4}}{4m^{2}}|\mathbf{k'}|^{2}}}\frac{|\mathbf{k}|}{|\mathbf{k'}|}.\label{gapeq2}
\end{eqnarray}
The second term in the right hand side is logarithmically divergent and dominates the whole right part of (\ref{gapeq2}). Using a hard cutoff $\Lambda$, the leading contribution to $\Delta$ is approximately independent of $|\mathbf{k}|$:
\begin{equation}
\Delta\approx\beta^{2}\Lambda e^{-\frac{2\pi}{\gamma}}.
\end{equation}
In the last expression the parameter $\gamma$ is controlled by the value of $\alpha_{r}$, so when $\epsilon_{r}\sim1$, $\gamma\sim\alpha\approx1/137$ (now the most unfavorable situation) the exponential is extremely small, meaning that the lower bound for $T_{c}$ is nearly zero. However, because we are working with a long wavelength effective theory the dominant contribution to the superconducting order parameter $\Delta_{\mathbf{k}}$ comes from small energies around the Fermi level, implying that a reliable estimation for $T_{c}$ is the one found in the small momentum regime. We have to mention that the p-wave symmetry of the order parameter is consistent with two basic facts: as we said, the time reversal symmetry is already broken because the magnetic coating and because the spin polarization of all electrons is the same, determined by the sign of $m$, so the origin of the p-wave symmetry of $\Delta_{\mathbf{k}}$ and the antisymmetry of the Cooper pair is purely orbital.

Because of the p-wave symmetry of the superconducting order parameter and the dimensionality of the problem, we can speculate with the fact that the superconducting transition will be of Kosterlitz-Thouless (KT) type, in a similar way to the transition found in superconducting thin films\cite{Doniach79,Halperin79}. The possible existence of vortices in $\Delta(\mathbf{r})$ might imply the existence of Majorana fermion states at the core of such vortices, according to ref.\cite{Read00}. This opens another route for searching Majorana zero modes in topological insulators.

\emph{Conclusions.} It is found that the topological magnetoelectric term present in a three dimensional topological insulator strongly modifies the electron-electron interaction when it becomes observable. The Induced Chern-Simons term gives rise to an attractive term that eventually induces a superconducting instability in the surface electronic states when the standard Coulomb repulsion is (statically) screened. The superconducting order parameter enjoys p-wave symmetry, which is consistent with the fact that time reversal symmetry is broken and the antisymmetry of the pair wavefunction is not determined by the spin part of the electronic wavefunctions but it has orbital origin. We have found an upper bound for the transition temperature $T_{c}$ to be of the order of $6$K. Finally, we have commented about the possibility of KT character of the transition and the existence of Majorana zero modes at the core of the vortices appearing in $\Delta(\mathbf{r})$.

\emph{Acknowledgments.} The author gratefully acknowledges conversations with A. G. Grushin, B. Valenzuela and M. A. H. Vozmediano. The author also acknowledges financial support from EPRSC Science and Innovation Award EP/G035954.


\begin{thebibliography}{23}
\expandafter\ifx\csname natexlab\endcsname\relax\def\natexlab#1{#1}\fi
\expandafter\ifx\csname bibnamefont\endcsname\relax
  \def\bibnamefont#1{#1}\fi
\expandafter\ifx\csname bibfnamefont\endcsname\relax
  \def\bibfnamefont#1{#1}\fi
\expandafter\ifx\csname citenamefont\endcsname\relax
  \def\citenamefont#1{#1}\fi
\expandafter\ifx\csname url\endcsname\relax
  \def\url#1{\texttt{#1}}\fi
\expandafter\ifx\csname urlprefix\endcsname\relax\def\urlprefix{URL }\fi
\providecommand{\bibinfo}[2]{#2}
\providecommand{\eprint}[2][]{\url{#2}}

\bibitem[{\citenamefont{Hsieh et~al.}(2008)}]{Hsieh08}
\bibinfo{author}{\bibfnamefont{D.}~\bibnamefont{Hsieh}} \bibnamefont{et~al.},
  \bibinfo{journal}{Nature} \textbf{\bibinfo{volume}{452}},
  \bibinfo{pages}{970} (\bibinfo{year}{2008}).

\bibitem[{\citenamefont{Hsieh et~al.}(2009)}]{Hsieh09}
\bibinfo{author}{\bibfnamefont{D.}~\bibnamefont{Hsieh}} \bibnamefont{et~al.},
  \bibinfo{journal}{Science} \textbf{\bibinfo{volume}{323}},
  \bibinfo{pages}{919} (\bibinfo{year}{2009}).

\bibitem[{\citenamefont{Xia et~al.}(2009)}]{Xia09}
\bibinfo{author}{\bibfnamefont{Y.}~\bibnamefont{Xia}} \bibnamefont{et~al.},
  \bibinfo{journal}{Nat. Phys.} \textbf{\bibinfo{volume}{5}},
  \bibinfo{pages}{398} (\bibinfo{year}{2009}).

\bibitem[{\citenamefont{Qi et~al.}(2008)\citenamefont{Qi, Hughes, and
  Zhang}}]{Qi08}
\bibinfo{author}{\bibfnamefont{X.~L.} \bibnamefont{Qi}},
  \bibinfo{author}{\bibfnamefont{T.~L.} \bibnamefont{Hughes}},
  \bibnamefont{and} \bibinfo{author}{\bibfnamefont{S.-C.} \bibnamefont{Zhang}},
  \bibinfo{journal}{Phys. Rev. B} \textbf{\bibinfo{volume}{78}},
  \bibinfo{pages}{195424} (\bibinfo{year}{2008}).

\bibitem[{\citenamefont{Wilczek}(1987)}]{Wilczek87}
\bibinfo{author}{\bibfnamefont{F.}~\bibnamefont{Wilczek}},
  \bibinfo{journal}{Phys. Rev. Lett.} \textbf{\bibinfo{volume}{58}},
  \bibinfo{pages}{1799} (\bibinfo{year}{1987}).

\bibitem[{\citenamefont{Fu and Kane}(2008)}]{Fu08}
\bibinfo{author}{\bibfnamefont{L.}~\bibnamefont{Fu}} \bibnamefont{and}
  \bibinfo{author}{\bibfnamefont{C.~L.} \bibnamefont{Kane}},
  \bibinfo{journal}{Phys. Rev. Lett.} \textbf{\bibinfo{volume}{100}},
  \bibinfo{pages}{096407} (\bibinfo{year}{2008}).

\bibitem[{\citenamefont{Grushin and Cortijo}(2011)}]{Grushin11}
\bibinfo{author}{\bibfnamefont{A.~G.} \bibnamefont{Grushin}} \bibnamefont{and}
  \bibinfo{author}{\bibfnamefont{A.}~\bibnamefont{Cortijo}},
  \bibinfo{journal}{Phys. Rev. Lett.} \textbf{\bibinfo{volume}{106}},
  \bibinfo{pages}{020403} (\bibinfo{year}{2011}).

\bibitem[{\citenamefont{Hasan and Kane}(2010)}]{Kane10}
\bibinfo{author}{\bibfnamefont{M.~Z.} \bibnamefont{Hasan}} \bibnamefont{and}
  \bibinfo{author}{\bibfnamefont{C.~L.} \bibnamefont{Kane}},
  \bibinfo{journal}{Rev. Mod. Phys.} \textbf{\bibinfo{volume}{82}},
  \bibinfo{pages}{3045} (\bibinfo{year}{2010}).

\bibitem[{\citenamefont{Santos et~al.}(2010)\citenamefont{Santos, Neupert,
  Chamon, and Mudry}}]{Santos10}
\bibinfo{author}{\bibfnamefont{L.}~\bibnamefont{Santos}},
  \bibinfo{author}{\bibfnamefont{T.}~\bibnamefont{Neupert}},
  \bibinfo{author}{\bibfnamefont{C.}~\bibnamefont{Chamon}}, \bibnamefont{and}
  \bibinfo{author}{\bibfnamefont{C.}~\bibnamefont{Mudry}},
  \bibinfo{journal}{Phys. Rev. B} \textbf{\bibinfo{volume}{81}},
  \bibinfo{pages}{184502} (\bibinfo{year}{2010}).

\bibitem[{\citenamefont{Fu and Berg}(2010)}]{Fu10}
\bibinfo{author}{\bibfnamefont{L.}~\bibnamefont{Fu}} \bibnamefont{and}
  \bibinfo{author}{\bibfnamefont{E.}~\bibnamefont{Berg}},
  \bibinfo{journal}{Phys. Rev. Lett.} \textbf{\bibinfo{volume}{105}},
  \bibinfo{pages}{097001} (\bibinfo{year}{2010}).

\bibitem[{\citenamefont{Hor et~al.}(2010)}]{Hor10}
\bibinfo{author}{\bibfnamefont{Y.~S.} \bibnamefont{Hor}} \bibnamefont{et~al.},
  \bibinfo{journal}{Phys. Rev. Lett.} \textbf{\bibinfo{volume}{104}},
  \bibinfo{pages}{057001} (\bibinfo{year}{2010}).

\bibitem[{\citenamefont{Wray et~al.}(2010)}]{Wray10}
\bibinfo{author}{\bibfnamefont{L.~A.} \bibnamefont{Wray}} \bibnamefont{et~al.},
  \bibinfo{journal}{Nat. Phys.} \textbf{\bibinfo{volume}{6}},
  \bibinfo{pages}{855} (\bibinfo{year}{2010}).

\bibitem[{\citenamefont{Greiter et~al.}(1992)\citenamefont{Greiter, Wen, and
  Wilczek}}]{Greiter92}
\bibinfo{author}{\bibfnamefont{M.}~\bibnamefont{Greiter}},
  \bibinfo{author}{\bibfnamefont{X.~G.} \bibnamefont{Wen}}, \bibnamefont{and}
  \bibinfo{author}{\bibfnamefont{F.}~\bibnamefont{Wilczek}},
  \bibinfo{journal}{Nucl. Phys. B} \textbf{\bibinfo{volume}{374}},
  \bibinfo{pages}{567} (\bibinfo{year}{1992}).

\bibitem[{\citenamefont{Ivanov}(2001)}]{Ivanov01}
\bibinfo{author}{\bibfnamefont{D.~A.} \bibnamefont{Ivanov}},
  \bibinfo{journal}{Phys. Rev. Lett.} \textbf{\bibinfo{volume}{86}},
  \bibinfo{pages}{268} (\bibinfo{year}{2001}).

\bibitem[{\citenamefont{Kogan}(1989)}]{Kogan89}
\bibinfo{author}{\bibfnamefont{Y.~I.} \bibnamefont{Kogan}},
  \bibinfo{journal}{JETP Lett.} \textbf{\bibinfo{volume}{49}},
  \bibinfo{pages}{225} (\bibinfo{year}{1989}).

\bibitem[{\citenamefont{Girotti et~al.}(1992)\citenamefont{Girotti, Gomes, and
  da~Silva}}]{Girotti92}
\bibinfo{author}{\bibfnamefont{H.~O.} \bibnamefont{Girotti}},
  \bibinfo{author}{\bibfnamefont{M.}~\bibnamefont{Gomes}}, \bibnamefont{and}
  \bibinfo{author}{\bibfnamefont{A.~J.} \bibnamefont{da~Silva}},
  \bibinfo{journal}{Phys. Lett. B} \textbf{\bibinfo{volume}{274}},
  \bibinfo{pages}{357} (\bibinfo{year}{1992}).

\bibitem[{\citenamefont{Zhang. et~al.}(2009)}]{Zhang09}
\bibinfo{author}{\bibfnamefont{H.}~\bibnamefont{Zhang.}} \bibnamefont{et~al.},
  \bibinfo{journal}{Nat. Phys.} \textbf{\bibinfo{volume}{5}},
  \bibinfo{pages}{438} (\bibinfo{year}{2009}).

\bibitem[{\citenamefont{Chen et~al.}(2010)}]{Chen10}
\bibinfo{author}{\bibfnamefont{Y.~L.} \bibnamefont{Chen}} \bibnamefont{et~al.},
  \bibinfo{journal}{Science} \textbf{\bibinfo{volume}{329}},
  \bibinfo{pages}{659} (\bibinfo{year}{2010}).

\bibitem[{\citenamefont{Gonzalez et~al.}(1994)\citenamefont{Gonzalez, Guinea,
  and Vozmediano}}]{Gonzalez94}
\bibinfo{author}{\bibfnamefont{J.}~\bibnamefont{Gonzalez}},
  \bibinfo{author}{\bibfnamefont{F.}~\bibnamefont{Guinea}}, \bibnamefont{and}
  \bibinfo{author}{\bibfnamefont{M.~A.~H.} \bibnamefont{Vozmediano}},
  \bibinfo{journal}{Nucl. Phys. B} \textbf{\bibinfo{volume}{424}},
  \bibinfo{pages}{595} (\bibinfo{year}{1994}).

\bibitem[{\citenamefont{Chaichian et~al.}(1998)\citenamefont{Chaichian, Chen,
  and Fainberg}}]{Chaichian98}
\bibinfo{author}{\bibfnamefont{M.}~\bibnamefont{Chaichian}},
  \bibinfo{author}{\bibfnamefont{W.~F.} \bibnamefont{Chen}}, \bibnamefont{and}
  \bibinfo{author}{\bibfnamefont{V.~Y.} \bibnamefont{Fainberg}},
  \bibinfo{journal}{Eur. Phys. J. C} \textbf{\bibinfo{volume}{5}},
  \bibinfo{pages}{545} (\bibinfo{year}{1998}).

\bibitem[{\citenamefont{Doniach and Huberman}(1979)}]{Doniach79}
\bibinfo{author}{\bibfnamefont{S.}~\bibnamefont{Doniach}} \bibnamefont{and}
  \bibinfo{author}{\bibfnamefont{B.~A.} \bibnamefont{Huberman}},
  \bibinfo{journal}{Phys. Rev. Lett.} \textbf{\bibinfo{volume}{42}},
  \bibinfo{pages}{1169} (\bibinfo{year}{1979}).

\bibitem[{\citenamefont{Halperin and Nelson}(1979)}]{Halperin79}
\bibinfo{author}{\bibfnamefont{B.~I.} \bibnamefont{Halperin}} \bibnamefont{and}
  \bibinfo{author}{\bibfnamefont{D.~R.} \bibnamefont{Nelson}},
  \bibinfo{journal}{J. Low Temp. Phys} \textbf{\bibinfo{volume}{36}},
  \bibinfo{pages}{599} (\bibinfo{year}{1979}).

\bibitem[{\citenamefont{Read and Green}(2000)}]{Read00}
\bibinfo{author}{\bibfnamefont{N.}~\bibnamefont{Read}} \bibnamefont{and}
  \bibinfo{author}{\bibfnamefont{D.}~\bibnamefont{Green}},
  \bibinfo{journal}{Phys. Rev. B} \textbf{\bibinfo{volume}{61}},
  \bibinfo{pages}{10267} (\bibinfo{year}{2000}).

\end{thebibliography}

\end{document}